\documentclass[runningheads]{llncs}
\usepackage{multirow}
\usepackage{pgfplots}
\usepackage{dblfloatfix}
\usepackage{amsmath}
\usepackage{amsfonts}
\usepackage{breakcites}
\usepackage{graphicx}
\begin{document}
\title{Item Cold Start Recommendation via Adversarial Variational Autoencoder Warm-up}
\titlerunning{Item Cold Start Recommendation via AVAEW}
\author{Shenzheng Zhang, Qi Tan, Xinzhi Zheng, Yi Ren, Xu Zhao, Wang Nian} 
\institute{Tencent News\\ 
$\spadesuit$ Corresponding author: \email{qjzcyzhang@tencent.com}}

\maketitle

\begin{abstract}
With numerous pieces of information emerging daily and greatly influencing people's lives, large-scale recommendation systems are necessary for timely bridging the users with their desired information. However, the existing widely used embedding-based recommendation systems have a shortcoming in recommending new items because little interaction data is available for training new item ID embedding, which is recognized as item cold start problem. The gap between the randomly initialized item ID embedding and the well-trained warm item ID embedding makes the cold items hard to suit the recommendation system, which is trained on the data of historical warm items. To alleviate the performance decline of new items recommendation, the distribution of the new item ID embedding should be close to that of the historical warm items. To achieve this goal, we propose an Adversarial Variational Autoencoder Warm-up model (AVAEW) to generate warm-up item ID embedding for cold items. Specifically, we develop a conditional variational autoencoder model to leverage the side information of items for generating the warm-up item ID embedding. Particularly, we introduce an adversarial module to enforce the alignment between warm-up item ID embedding distribution and historical item ID embedding distribution. We demonstrate the effectiveness and compatibility of the proposed method by extensive offline experiments on public datasets and online A/B tests on a real-world large-scale news recommendation platform.
\end{abstract}


\section{Introduction}
Recommendation system has become more important in the era of information, where lots of information content emerge each day and play an important role in each person's life~\cite{qi2021pp,cheng2016music,wang2020off,zhao2019recommending}. In order to better capture the characteristics of recommended content, recommendation systems usually learn an item ID embedding from the user-item interaction data~\cite{guo2017deepfm,cheng2016wide,wang2017deep}.
However, such systems suffer from the item cold start problem that the item ID embedding is not trained sufficiently because little interaction information has been gained~\cite{vartak2017meta,volkovs2017dropoutnet,wei2021contrastive,Ou2021Learning}. Without sufficient training data for cold items, the distribution of the cold item ID embedding is significantly apart from that of the warm item ID embedding. 
Since the warm item ID contributes most of the samples in training the recommendation system, such embedding distribution gap makes the cold item hard to suit the recommendation system.

The existing methods proposed to solve the item cold start problem mainly fall into three categories: 1) promoting the robustness of the recommendation model in absence of item ID embedding, such as using dropout or masking on item ID embedding in the model training~\cite{volkovs2017dropoutnet,zheng2021multi,shiattention2018}; 2) improving the learning efficiency with a limited amount of interaction data, such as using meta-learning approaches to quickly adapt the recommendation system to new item~\cite{vartak2017meta,pan2019warm,sunform2021,fengcmml2021}; and 3) leveraging the side information of items to facilitate the initialization of the item ID embedding~\cite{wei2021contrastive,zhu2021learning,zhu2020recommendation,chen2022generative}.

However, the methods in the first category cannot fully train the item ID embedding and utilize this collaborative information, and thus have deficient performance in adapting to a limited number of user-item interactions in the warm-up phases. The methods in the second category alleviate the inefficient training problem but start with randomly initialized item embedding, which may be quite different from the well-trained hot item embeddings. Such distribution gap slows down the conversion of cold items to warm-up items. The methods in the third category provide a better initialization for the conversion, but none of the existing methods rigorously consider the embedding distribution gap problem and have no strict guarantee of alleviating the distribution gap in the warm-up process. Recently, some approaches have attempted to solve the problem of distribution gap, e.g., GAR~\cite{chen2022generative} designs an adversarial training strategy: train an
item generator via maximizing the ranking scores of cold items and train the
recommendation model adversarially to decrease the corresponding ranking scores. They are closely related to our work. However, they do not directly consider the consistency of the distribution, nor do they strictly guarantee to reduce the distribution gap during the warm-up process, which reduces their recommendation effect.

To address the problem of item ID embedding gap in item cold start, our core idea in this paper is to generate a better warm-up item ID embedding whose distribution is close to one of the warm items ID embedding by introducing an adversarial model to decrease the embedding distribution gap. Particularly, we proposed a method named Adversarial Variational AutoEncoder Warm-up to address the item cold start problem. AVAEW leverages the side information of items to generate the warm-up item ID embedding that is similar to the distribution of historical warm item embedding. We further introduce an adversarial module into the encoder-decoder generative model to enforce the alignment between warm item ID embedding distribution and historical item embedding. Since our methods only act on item ID embedding, our method is compatible with different kinds of recommendation backbone models. Moreover, as the proposed method only utilizes the side information of the items which is available in most real-world scenarios, it can be easily applied in large-scale industrial systems without further development of data or training pipeline, such as item-item relationship required by graph embedding learning methods~\cite{LeeMeLU2019,Ou2021Learning} or separation of global and local update in meta-learning method~\cite{sunform2021}. We evaluate the proposed method by extensive offline experiments on public datasets and online A/B tests on a large-scale real-world recommendation platform.

The contributions in this paper can be summarized as follow:
\begin{enumerate}
\item We first directly and rigorously consider the item ID embedding gap problem in the item cold start recommendation and propose AVAEW to alleviate the gap in an adversarial manner. AVAEW is able to ensures the consistent distribution of generated results.
\item AVAEW has no extra data requirements, which
makes it easy to be deployed in the online scenario and compatible in equipping with different embedding-based recommendation models. 
\item We conduct extensive offline and online experiments to demonstrate the
effectiveness of AVAEW compared with other state-of-art warm-up methods
and compatibility in equipping with different recommendation models.
\end{enumerate}

\section{Proposed Method}

\subsection{Problem Formulation}

\textbf{Click-Through Rate (CTR)} In this work, we focus on the Click-Through Rate (CTR) prediction task, one of the most common tasks in recommendation systems~\cite{zhou2018deep,guo2017deepfm}.
For a user-item pair, the CTR model aims to generate a prediction $y$ with inputs of item-related features $\mathcal{V}$ and other features $\mathcal{U}$ (such as user profile data and context data). The model is trained by minimizing the discriminative loss between the model prediction and corresponding interaction observation $o$ (i.e., click or unclick). The Binary Cross Entropy loss is commonly adopted as the discriminative loss:
\begin{equation}
   \mathcal{L}_{M} = \mathbb{E} [- o \log(y) - (1-o) \log (1-y)].
\end{equation}

\textbf{Item Cold Start Problem} Following the embedding learning framework~\cite{vasile2016meta,grbovic2018real,wang2018billion}, recommendation models learn the embeddings for the discrete input data, among which the item ID embedding is one of the most critical representations for the items. We partition the item side features into two parts: the item ID embedding $(\mathbf{v}_\mathcal{I}$ and side information $\mathcal{V}_S$). Then the recommendation model can be formulated as:
\begin{equation}
    y = f(\mathbf{v}_\mathcal{I}, \mathcal{V}_\mathcal{S}, \mathcal{U}; \theta)
    \label{EQU:prediction}
\end{equation}
where $\theta$ is the set of parameters of the recommendation model.
However, the item ID embedding of the new items cannot be learned sufficiently due to the lack of enough training interaction data, causing the distribution gap with that of warm items and thus the decline of recommendation performance. Therefore, we need to use the limited amount of new items' interaction data effectively to quickly adapt their item ID embeddings, which is regarded as the item cold start problem, and the new items are called cold items in the recommendation system. In this work, we aim to alleviate this problem by generating better item ID embeddings for these new cold items such that their distribution is close to the item ID embedding distribution of the warm items.

\subsection{Adversarial Variational AutoEncoder Warm-up}
\begin{figure}[t]
    \centering
    \includegraphics[width=1\linewidth]{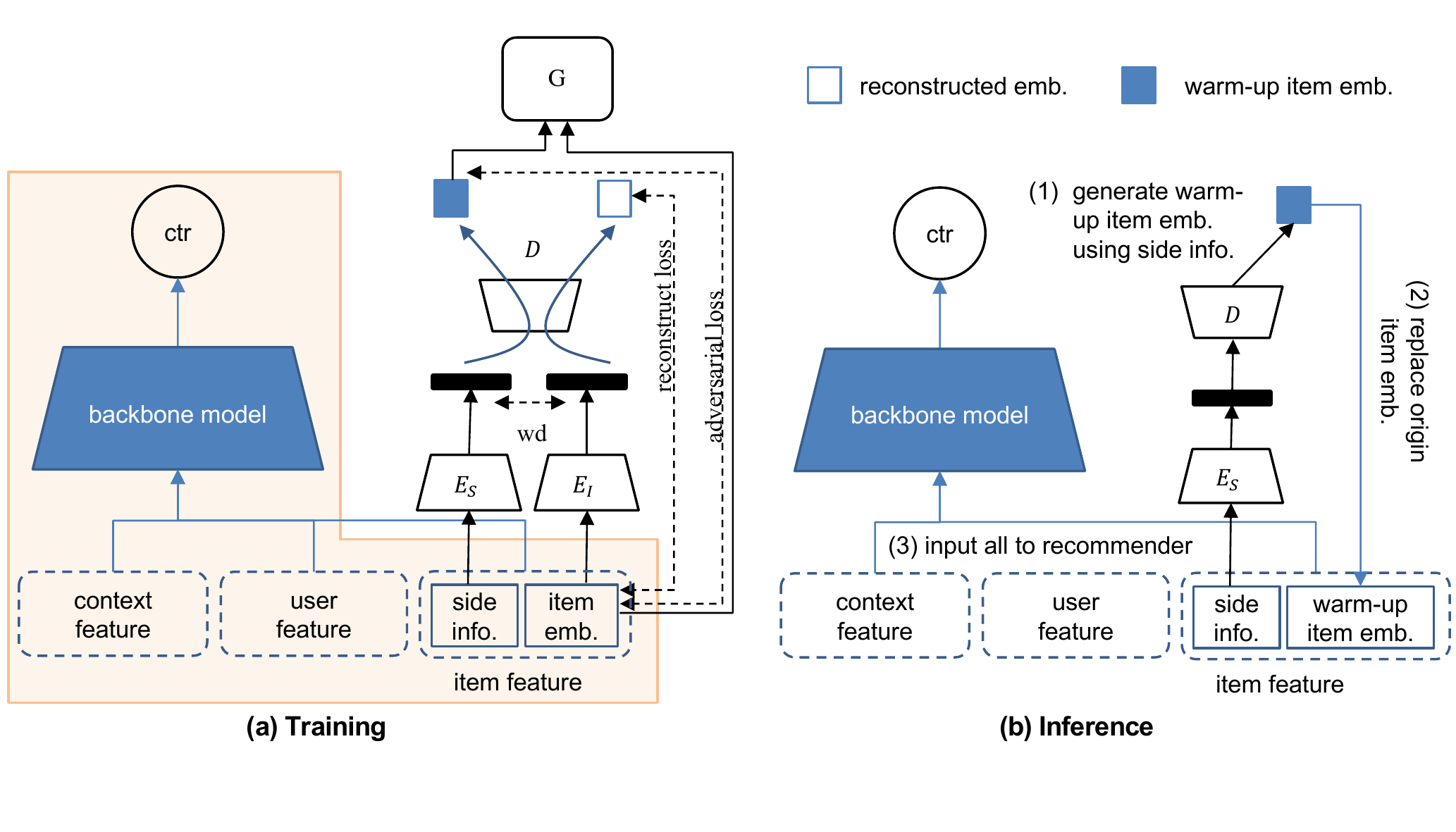}
    \caption{Framework of proposed Adversarial Variational Auto-Encoder Warm-up model, which mainly consists of four components additional to the backbone model (shown in the orange box): 1) the ID embedding encoder $E_I$ generates the latent space representation of the item ID embedding; 2) the prior encoder $E_S$ utilizes the side information of item to generate the customized prior distribution for warm-up item ID embedding; 3) the decoder $D$ generates the reconstructed/warm-up item ID embedding from the latent representation; 4) the adversarial module discriminates the warm-up embedding and real embedding to align their distributions. In the inference phase, we 1) generate the warm-up item embedding for the cold items; 2) use the warm-up item embedding as the item embedding, and 3) inputs all user, context, and item features into the recommendation model for prediction.}
    \label{fig:framework}
    \vspace{-3mm}
\end{figure}

We introduce the proposed Adversarial Variational AutoEncoder Warm-up method in this part. Figure~\ref{fig:framework} shows the framework of our proposed model. We develop a VAE-based latent space model to generate a warm-up item ID embedding for the cold items. In order to close the ID embedding distribution gap between the cold and warm items, we use an adversarial module to align the warm-up item ID embedding distribution with the historical warm item ID embedding.\\

\textbf{Item ID Embedding Encoder \& Decoder}
We develop the item ID embedding encoder \& decoder following the VAE framework~\cite{walker2016uncertain,sohn2015learning}. The item ID encoder module takes the item ID embedding as input and generates the corresponding latent distribution. The decoder samples a latent variable from the generated latent distribution and reconstructs the item ID embedding based on the sampled variable. The encoder-decoder module can be formulated as:
\begin{equation}
\begin{aligned}
      \mathbf{\mu},\sigma &= g_{I}(\mathbf{v}_\mathcal{I};\theta_{I}) \\
      \mathbf{z} \sim \mathbb{N}(\mu, &\Sigma), \mathrm{diag}(\Sigma)=\sigma \\
      \hat{\mathbf{v}}_\mathcal{I} &= g_{D}(\mathbf{z};\theta_{D})\\
\end{aligned}
\end{equation}
where $\mu$ and $\sigma$ are k-dimension vectors, $\theta_{I}$ and $\theta_{D}$ are the set of parameters of item ID encoder $E_I$ and encoder $D$, respectively. We train the encoder-decoder module via the reconstruction loss:
\begin{equation}
    \mathcal{L}_{R} = ||\mathbf{v}_\mathcal{I} -  \hat{\mathbf{v}}_\mathcal{I}||_2^2 
\end{equation}

\textbf{Prior Encoder}
Due to the deficiency of training data, the item ID embeddings of cold items are nearly randomly initialized.  
To provide a more informative item ID embedding, the prior encoder $E_s$ takes the side information of the items as inputs and generated the warm-up item ID embedding from the prior distribution
\begin{equation}
\begin{aligned}
      \mu_p,\sigma_p &= g_{S}(\mathcal{V}_\mathcal{S};\theta_{S}). \\
      \mathbf{z}_p &\sim \mathbb{N}(\mu_p, \Sigma_p), \mathrm{diag}(\Sigma_p)=\sigma_p \\
    \bar{\mathbf{v}}_\mathcal{I} & = g_{D}(\mathbf{z}_p; \theta_D)
\end{aligned}
\end{equation}

We minimize the Wasserstein Distance~\cite{shen2018wasserstein} between the encoded distribution and prior distribution to train the prior encoder: \\
\begin{equation}
    \mathcal{L}_{WD} = WD(\mathbb{N}(\mu, \Sigma), \mathbb{N}(\mu_p, \Sigma_p)).
\end{equation}
In addition, we replace the warm-up item embedding in Equation~\ref{EQU:prediction} to generate the output and calculate the recommendation CTR loss:
\begin{equation}
    \begin{aligned}
          \bar{y} & = f( \bar{\mathbf{v}}_\mathcal{I}, \mathcal{V}_\mathcal{S}, \mathcal{U}; \theta)\\
         \mathcal{L}_{CTR} & = \mathbb{E} [- o \log(\bar{y}) - (1-o) \log (1-\bar{y})].
    \end{aligned}
\end{equation}
\textbf{Adversarial Alignment}
In the inference stage, $\bar{\mathbf{v}}_\mathcal{I}$ is used as the item ID embedding for the cold item. Though we used reconstruction loss between ${\mathbf{v}}_\mathcal{I}$ and $\hat{\mathbf{v}}_\mathcal{I}$ and Wasserstein Distance between the encoded distribution and prior distribution in training different parts of the modules, we don't have direct regulation to alleviate the gap between warm-up item ID embedding $\bar{\mathbf{v}}_\mathcal{I}$ and the origin item ID embedding ${\mathbf{v}}_\mathcal{I}$, which impairs the performance of the recommendation model.

To directly address this problem, we propose to introduce an extra regulation using the adversarial network to enforce the closeness of warm-up item ID embedding and the origin item ID embedding. 
A classifier $G$ is trained to discriminate the origin item ID embedding and warm-up ID embedding by minimizing the loss:
\begin{equation}
    \mathcal{L}_D = -\mathbb{E}_{{\mathbf{v}}_\mathcal{I}\sim P_{{\mathbf{v}}_\mathcal{I}}}[\log G({\mathbf{v}}_\mathcal{I})] - \mathbb{E}_{\mathbf{z}_p \sim \mathbb{N}(\mu_p,\sigma_p)}[\log (1 - G(g_D(\mathbf{z}_p)))].
\end{equation}

To train the encoders, we use group mean feature matching~\cite{bao2017cvae}, which closes the empirical mean of the prior distribution and encoded distribution:
\begin{equation}
    \mathcal{L}_{GD} = ||\mathbb{E}_{{\mathbf{v}}_\mathcal{I}\sim P_{{\mathbf{v}}_\mathcal{I}}}[f_G(\mathbf{v}_\mathcal{I})] - \mathbb{E}_{\mathbf{z}_p \sim \mathbb{N}(\mu_p,\sigma_p)}[f_G(g_D(\mathbf{z}_p)]||^2_2,
\end{equation}
and pairwise feature matching, which closes the pairwise sample distance:
\begin{equation}
    \mathcal{L}^{'}_{GD} = \mathbb{E}_{{\mathbf{v}}_\mathcal{I}\sim P_{{\mathbf{v}}_\mathcal{I}}, \mathbf{z}_p \sim \mathbb{N}(\mu_p,\sigma_p)} ||f_G(\mathbf{v}_\mathcal{I}) - f_G(g_D(\mathbf{z}_p))||^2_2,
\end{equation}
where $f_G(x)$ is the features on an intermediate layer of the classifier. We choose the last layer of the classifier as the intermediate layer.\\

\textbf{Model Training}
We train the backbone model and AVAEW warm-up parts consecutively,
We first train the recommendation model via minimizing $\mathcal{L}_{M}$. Then we train the warm-up item ID generative module ($E_I$, $E_S$, $D$) and (2) classifier $G$, alternatively.
The warm-up item ID embedding generative module is trained via minimizing the total loss:
\begin{equation}
    \mathcal{L}_{w} = \underbrace{\mathcal{L}_{CTR} + \alpha \mathcal{L}_{R} + \beta \mathcal{L}_{WD}}_{\mathrm{variational}} + \underbrace{\xi \mathcal{L}_{GD} + \xi' \mathcal{L}^{'}_{GD}}_{\mathrm{adversarial}},
    \label{EQU:LOSS_FUNC}
\end{equation}
where $\alpha, \beta, \xi, \xi'$ are the hyper-parameters. The classifier $G$ is trained via minimizing $\mathcal{L}_D$.

\section{Validation}
\subsection{Offline Validation}
We first validate the effectiveness of the proposed AVAEW offline on the representative public datasets.\\

\textbf{Experiment Setting} We conduct experiments on the following three public datasets:  MovieLens-1M~\footnote{\url{https://grouplens.org/datasets/movielens/1m/}}, MovieLens-25M~\footnote{\url{https://grouplens.org/datasets/movielens/25m/}}, and Taobao Display Ad Click~\footnote{\url{https://tianchi.aliyun.com/dataset/dataDetail?dataId=56}}. A detailed description of these datasets can be found in the Supplementary Material.
    
To validate the recommendation performance of the proposed method in different phases, we preprocess the datasets following \cite{zhao2022improving,pan2019warm,zhu2021learning}. Specifically, we label items as old or new based on their frequency, where items with more than $N$ labeled instances are old and others are new. 
We set $N$ of 200, 2700, and 2000 for Movielens-1M, Movielens-25M, and Taobao Ad data.
Then the new item instances are divided into four sets (denoted as warm-a, -b, -c, and test) according to the order of their release timestamp. Specifically, for each new item, the first $K$ ratings are divided into warm-a, second $K$ ratings are warm-b, third $K$ ratings are warm-c and the later ratings are test set. The new items with less than $3K$ ratings are discarded. We set $K$ to 20, 40, and 500.
By setting $N$ and $K$ to these values, the ratio between the number of news items and old items is around 8:2, which approximates the distribution of long-tail items. The ratio between the sizes of warm-a, -b, -c, and test sets is approximately 1:1:1:3.\\

\textbf{Backbones and Baselines}
As summarized in Section 1, there are three general categories for solving the item cold start problem.
We choose the State-Of-The-Art (SOTA) methods in each category for comparison: 
\begin{itemize}
    \item DropoutNet~\cite{volkovs2017dropoutnet} improves the robustness by applying dropout technology to mitigate the model’s dependency on item ID. 
    \item Meta embedding (Meta-E)~\cite{pan2019warm} learns the ID embedding for new items fast adaption with meta-learning technology.
    \item MWUF~\cite{zhu2021learning} and CVAR~\cite{zhao2022improving} use side information of items to generate warm-up item ID embedding by meta Scaling/Shifting networks and conditional variational auto-encoder, respectively.  GAR~\cite{chen2022generative} attempts to increase the ranking score of cold items by training a warm-up item embedding generator in an adversarial manner. 
\end{itemize}

Because AVAEW only acts on the item ID embedding, it can be equipped with different kinds of embedding-based backbone recommendation models. To demonstrate the compatibility, conduct experiments upon the following representative backbones that are widely used in practical recommendation: Factorization Machine (FM)~\cite{steffen2010factorization}, DeepFM~\cite{guo2017deepfm}, Wide\&Deep~\cite{cheng2016wide}, Deep \& Cross Network (DCN)~\cite{wang2017deep}, IPNN~\cite{qu2016product} and OPNN~\cite{qu2016product}.

\begin{table}[h]
\def\arraystretch{1.1}
\setlength\tabcolsep{3pt}
\small
\begin{tabular}{c|c|c|c|c|c|c|c}
\hline
\multirow{2}{*}{Datasets}      & \multirow{2}{*}{Methods} & \multicolumn{3}{c|}{DeepFM - backbone}                                                        & \multicolumn{3}{c}{IPNN - backbone}                                                          \\ \cline{3-8} 
                              &                          & \multicolumn{1}{c|}{warm-a}          & \multicolumn{1}{c|}{warm-b}          & warm-c          & \multicolumn{1}{c|}{warm-a}          & \multicolumn{1}{c|}{warm-b}          & warm-c          \\ \hline
\multirow{6}{*}{MovieLens-1M} & base            & \multicolumn{1}{c|}{0.7452}          & \multicolumn{1}{c|}{0.7570}          & 0.7667          & \multicolumn{1}{c|}{0.7489}          & \multicolumn{1}{c|}{0.7609}          & 0.7715          \\ 

                              & DropoutNet               & \multicolumn{1}{c|}{0.7502}          & \multicolumn{1}{c|}{0.7590}          & 0.7667          & \multicolumn{1}{c|}{0.7404}          & \multicolumn{1}{c|}{0.7472}          & 0.7537          \\ 
                              & Meta-E                   & \multicolumn{1}{c|}{0.7439}          & \multicolumn{1}{c|}{0.7574}          & 0.7690          & \multicolumn{1}{c|}{0.7135}          & \multicolumn{1}{c|}{0.7309}          & 0.7460          \\ 
                              & MWUF                     & \multicolumn{1}{c|}{0.7451}          & \multicolumn{1}{c|}{0.7575}          & 0.7682          & \multicolumn{1}{c|}{0.7480}          & \multicolumn{1}{c|}{0.7614}          & 0.7719          \\ 
                              & CVAR                     & \multicolumn{1}{c|}{0.7848}          & \multicolumn{1}{c|}{0.7981}          & 0.8057          & \multicolumn{1}{c|}{0.7825}          & \multicolumn{1}{c|}{0.7970}          & 0.8022          \\ 
                              & GAR           & 0.7671          & 0.7776          & 0.7962          & 0.7837          & 0.7896          & 0.7991          \\
                              & AVAEW                    & \multicolumn{1}{c|}{\textbf{0.7909}} & \multicolumn{1}{c|}{\textbf{0.8025}} & \textbf{0.8064} & \multicolumn{1}{c|}{\textbf{0.7859}} & \multicolumn{1}{c|}{\textbf{0.7982}} & \textbf{0.8027} \\ \hline
\multirow{6}{*}{MovieLens-25M} & base            & \multicolumn{1}{c|}{0.7947}          & \multicolumn{1}{c|}{0.8040}          & 0.8098          & \multicolumn{1}{c|}{0.8065}          & \multicolumn{1}{c|}{0.8113}          & 0.8151          \\ 
                              & DropoutNet               & \multicolumn{1}{c|}{0.7936}          & \multicolumn{1}{c|}{0.8000}          & 0.8046          & \multicolumn{1}{c|}{0.8028}          & \multicolumn{1}{c|}{0.8053}          & 0.8075          \\ 
                              & Meta-E                   & \multicolumn{1}{c|}{0.8010}          & \multicolumn{1}{c|}{0.8072}          & 0.8110          & \multicolumn{1}{c|}{0.8094}          & \multicolumn{1}{c|}{0.8140}          & 0.8175          \\ 
                              & MWUF                     & \multicolumn{1}{c|}{0.7950}          & \multicolumn{1}{c|}{0.8046}          & 0.8106          & \multicolumn{1}{c|}{0.8058}          & \multicolumn{1}{c|}{0.8109}          & 0.8151          \\ 
                              & CVAR                     & \multicolumn{1}{c|}{0.8074}          & \multicolumn{1}{c|}{\textbf{0.8110}} & 0.8118          & \multicolumn{1}{c|}{0.8087}          & \multicolumn{1}{c|}{0.8189}          & 0.8217          \\ 
                              & GAR           & 0.7790          & 0.7841          & 0.7861          & 0.7974          & 0.7992          & 0.8005          \\
                              & AVAEW                    & \multicolumn{1}{c|}{\textbf{0.8079}} & \multicolumn{1}{c|}{0.8107}          & \textbf{0.8131} & \multicolumn{1}{c|}{\textbf{0.8184}} & \multicolumn{1}{c|}{\textbf{0.8226}} & \textbf{0.8249} \\ \hline
\multirow{6}{*}{Taobao-AD}    & base            & \multicolumn{1}{c|}{0.6127}          & \multicolumn{1}{c|}{0.6222}          & 0.6307          & \multicolumn{1}{c|}{0.6163}          & \multicolumn{1}{c|}{0.6230}          & 0.6291          \\ 
                              & DropoutNet               & \multicolumn{1}{c|}{0.6138}          & \multicolumn{1}{c|}{0.6227}          & 0.6308          & \multicolumn{1}{c|}{0.6162}          & \multicolumn{1}{c|}{0.6233}          & 0.6295          \\ 
                              & Meta-E                   & \multicolumn{1}{c|}{0.6035}          & \multicolumn{1}{c|}{0.6148}          & 0.6243          & \multicolumn{1}{c|}{0.6216}          & \multicolumn{1}{c|}{0.6287}          & 0.6350          \\ 
                              & MWUF                     & \multicolumn{1}{c|}{0.6079}          & \multicolumn{1}{c|}{0.6167}          & 0.6254          & \multicolumn{1}{c|}{0.6197}          & \multicolumn{1}{c|}{0.6260}          & 0.6319          \\ 
                              & CVAR                     & \multicolumn{1}{c|}{0.6140}          & \multicolumn{1}{c|}{0.6251}          & 0.6360          & \multicolumn{1}{c|}{0.6272}          & \multicolumn{1}{c|}{0.6347}          & 0.6413          \\ 
                              & GAR           & 0.6063          & 0.6112          & 0.6175          & 0.6149          & 0.6237          & 0.6314          \\
                              & AVAEW                    & \multicolumn{1}{c|}{\textbf{0.6186}} & \multicolumn{1}{c|}{\textbf{0.6306}} & \textbf{0.6384} & \multicolumn{1}{c|}{\textbf{0.6287}} & \multicolumn{1}{c|}{\textbf{0.6359}} & \textbf{0.6419} \\ \hline
\end{tabular}
\caption{Comparison result with other SOTA item warm-up methods. AUC metric of recommendation results are reported. The best performances are highlighted in bold.}
\label{TAB:COMP}
\vspace{-5mm}
\end{table}

\textbf{Implementation Details} For backbones models, the MLPs in backbones use the same structure with two dense layers (hidden units 16). 
The embedding size of each feature is fixed to 16. 
For a fair comparison, we use the same setting for all warm-up methods.
We use Adam as optimizer with learning rate set to 0.001 and mini-batch size set to 2048. We use old item instances to first pre-train the backbone model and train the AVAEW model with the fixed backbone model. Then we progressively feed warm-a, -b, -c data to train the backbone and AVAEW and evaluate the models on the test set. 
We take the AUC score as the evaluation metric.\\

\textbf{Validation Result}
First, we demonstrate the superiority of AVAEW compared with other SOTA warm-up methods. Table~\ref{TAB:COMP} shows the AUC of AVAEW and other SOTA warm-up methods with two backbones (DeepFM and IPNN) on three public datasets. As expected, as DropoutNet ignores the reliance on item ID embedding of the cold items to some extent, it cannot quickly fit the cold item with limited samples. DropoutNet performs similarly or even worse than the back backbone in the warm-c phase. Meta-E is designed to fast adaption but ignores the initial embedding in terms of performance. Thus when the warm-up data is very limited, for example after warm-a phase of MoiveLens-1M, the Meta-E cannot perform well. In the cases where the warm-up data is sufficient for overcoming this shortcomings of random initialized item embeddings, for example in warm-a phase of MovieLens-25M and Taobao-AD, Meta-E performs relatively well.
Compared with MWUF and CVAR, the biggest difference in AVAEW is that AVAEW adopts an adversarial module to merge the distribution of cold and hot item embeddings. 
The results show that AVAEW performs best in the warm phases, which reflects the benefits of quickly adapting the cold item ID embedding distribution by matching the warm item ID embedding distribution.
Due to the adoption of variational autoencoder, AVAEW alleviates the mode collapse problem compared to the pure GAN model~\cite{bao2017cvae}. Therefore, from the results, we can also see that AVAEW achieves better results than GAR.

\begin{table}[htpb]
\def\arraystretch{1.1}
\small
\begin{tabular}{cc|ccc|ccc|ccc}
\hline
 & \multirow{2}{*}{Methods} & \multicolumn{3}{c|}{MovieLens-1M}                                   & \multicolumn{3}{c|}{MovieLens-25M}                                 & \multicolumn{3}{c}{Taobao-AD}                                       \\ \cline{3-11} 
                           &                          & \multicolumn{1}{c|}{warm-a} & \multicolumn{1}{c|}{warm-b} & warm-c & \multicolumn{1}{c|}{warm-a} & \multicolumn{1}{c|}{warm-b} & warm-c & \multicolumn{1}{c|}{warm-a} & \multicolumn{1}{c|}{warm-b}  & warm-c  \\ \hline
\multirow{3}{*}{\rotatebox[origin=c]{90}{DeepFM}}    & Base                     & \multicolumn{1}{c|}{0.7452} & \multicolumn{1}{c|}{0.7570} & 0.7667 & \multicolumn{1}{c|}{0.7947} & \multicolumn{1}{c|}{0.8040} & 0.8098 & \multicolumn{1}{c|}{0.6127} & \multicolumn{1}{c|}{0.6222}  & 0.6307  \\ 
                           & AVAEW                    & \multicolumn{1}{c|}{0.7909} & \multicolumn{1}{c|}{0.8025} & 0.8064 & \multicolumn{1}{c|}{0.8077} & \multicolumn{1}{c|}{0.8122} & 0.8143 & \multicolumn{1}{c|}{0.6186} & \multicolumn{1}{c|}{0.6306}  & 0.6384  \\  
                           & improve               & \multicolumn{1}{c|}{6.12\%} & \multicolumn{1}{c|}{6.01\%} & 5.18\% & \multicolumn{1}{c|}{1.64\%} & \multicolumn{1}{c|}{1.02\%} & 0.56\% & \multicolumn{1}{c|}{0.96\%} & \multicolumn{1}{c|}{1.35\%}  & 1.23\%  \\ \hline
\multirow{3}{*}{\rotatebox[origin=c]{90}{WD}} & Base                     & \multicolumn{1}{c|}{0.7352} & \multicolumn{1}{c|}{0.7492} & 0.7608 & \multicolumn{1}{c|}{0.7860} & \multicolumn{1}{c|}{0.7926} & 0.7969 & \multicolumn{1}{c|}{0.5932} & \multicolumn{1}{c|}{0.6045}  & 0.6134  \\ 
                           & AVAEW                    & \multicolumn{1}{c|}{0.7764} & \multicolumn{1}{c|}{0.7946} & 0.7956 & \multicolumn{1}{c|}{0.8001} & \multicolumn{1}{c|}{0.8047} & 0.8059 & \multicolumn{1}{c|}{0.6126} & \multicolumn{1}{c|}{0.6212}  & 0.6317  \\ 
                           & improve               & \multicolumn{1}{c|}{5.60\%} & \multicolumn{1}{c|}{6.06\%} & 4.57\% & \multicolumn{1}{c|}{1.79\%} & \multicolumn{1}{c|}{1.53\%} & 1.13\% & \multicolumn{1}{c|}{3.27\%} & \multicolumn{1}{c|}{2.77\%}  & 2.99\%  \\ \hline
\multirow{3}{*}{\rotatebox[origin=c]{90}{FM}}        & Base                     & \multicolumn{1}{c|}{0.7411} & \multicolumn{1}{c|}{0.7513} & 0.7603 & \multicolumn{1}{c|}{0.7781} & \multicolumn{1}{c|}{0.7915} & 0.7980 & \multicolumn{1}{c|}{0.5909} & \multicolumn{1}{c|}{0.6135}  & 0.6278  \\ 
                           & AVAEW                    & \multicolumn{1}{c|}{0.7853} & \multicolumn{1}{c|}{0.7973} & 0.8031 & \multicolumn{1}{c|}{0.7936} & \multicolumn{1}{c|}{0.7973} & 0.8008 & \multicolumn{1}{c|}{0.6114} & \multicolumn{1}{c|}{0.6220}  & 0.6269  \\  
                           & improve             & \multicolumn{1}{c|}{5.96\%} & \multicolumn{1}{c|}{6.12\%} & 5.62\% & \multicolumn{1}{c|}{1.98\%} & \multicolumn{1}{c|}{0.73\%} & 0.35\% & \multicolumn{1}{c|}{3.47\%} & \multicolumn{1}{c|}{1.38\%}  & -0.13\% \\ \hline
\multirow{3}{*}{\rotatebox[origin=c]{90}{DCN}}       & Base                     & \multicolumn{1}{c|}{0.7251} & \multicolumn{1}{c|}{0.7400} & 0.7528 & \multicolumn{1}{c|}{0.7953} & \multicolumn{1}{c|}{0.8015} & 0.8058 & \multicolumn{1}{c|}{0.6210} & \multicolumn{1}{c|}{0.6304}  & 0.6377  \\  
                           & AVAEW                    & \multicolumn{1}{c|}{0.7839} & \multicolumn{1}{c|}{0.8023} & 0.8077 & \multicolumn{1}{c|}{0.8064} & \multicolumn{1}{c|}{0.8104} & 0.8123 & \multicolumn{1}{c|}{0.6309} & \multicolumn{1}{c|}{0.6375}  & 0.6431  \\ 
                           & improve               & \multicolumn{1}{c|}{8.11\%} & \multicolumn{1}{c|}{8.43\%} & 7.29\% & \multicolumn{1}{c|}{1.40\%} & \multicolumn{1}{c|}{1.11\%} & 0.80\% & \multicolumn{1}{c|}{1.59\%} & \multicolumn{1}{c|}{1.12\%}  & 0.85\%  \\ \hline
\multirow{3}{*}{\rotatebox[origin=c]{90}{OPNN}}      & Base                     & \multicolumn{1}{c|}{0.7432} & \multicolumn{1}{c|}{0.7579} & 0.7698 & \multicolumn{1}{c|}{0.8012} & \multicolumn{1}{c|}{0.8075} & 0.8120 & \multicolumn{1}{c|}{0.6257} & \multicolumn{1}{c|}{0.6323}  & 0.6384  \\ 
                           & AVAEW                    & \multicolumn{1}{c|}{0.7783} & \multicolumn{1}{c|}{0.7938} & 0.7997 & \multicolumn{1}{c|}{0.8136} & \multicolumn{1}{c|}{0.8168} & 0.8168 & \multicolumn{1}{c|}{0.6276} & \multicolumn{1}{c|}{0.6319}  & 0.6399  \\  
                           & improve              & \multicolumn{1}{c|}{4.73\%} & \multicolumn{1}{c|}{4.74\%} & 3.88\% & \multicolumn{1}{c|}{1.55\%} & \multicolumn{1}{c|}{1.15\%} & 0.60\% & \multicolumn{1}{c|}{0.30\%} & \multicolumn{1}{c|}{-0.07\%} & 0.23\%  \\ \hline
\multirow{3}{*}{\rotatebox[origin=c]{90}{IPNN}}      & Base                     & \multicolumn{1}{c|}{0.7489} & \multicolumn{1}{c|}{0.7609} & 0.7715 & \multicolumn{1}{c|}{0.8065} & \multicolumn{1}{c|}{0.8113} & 0.8151 & \multicolumn{1}{c|}{0.6239} & \multicolumn{1}{c|}{0.6311}  & 0.6377  \\  
                           & AVAEW                    & \multicolumn{1}{c|}{0.7859} & \multicolumn{1}{c|}{0.7982} & 0.8027 & \multicolumn{1}{c|}{0.8121} & \multicolumn{1}{c|}{0.8184} & 0.8202 & \multicolumn{1}{c|}{0.6287} & \multicolumn{1}{c|}{0.6359}  & 0.6419  \\ 
                           & improve              & \multicolumn{1}{c|}{4.95\%} & \multicolumn{1}{c|}{4.90\%} & 4.05\% & \multicolumn{1}{c|}{0.69\%} & \multicolumn{1}{c|}{0.87\%} & 0.62\% & \multicolumn{1}{c|}{0.77\%} & \multicolumn{1}{c|}{0.76\%}  & 0.67\%  \\ \hline
\end{tabular}
\caption{AUC of the base model and AVAEW on three datasets with different backbones.}
\label{TAB:DiffBackbones}
\vspace{-3.5mm}
\end{table}

\begin{figure}
    \centering
    \includegraphics[width=1\linewidth]{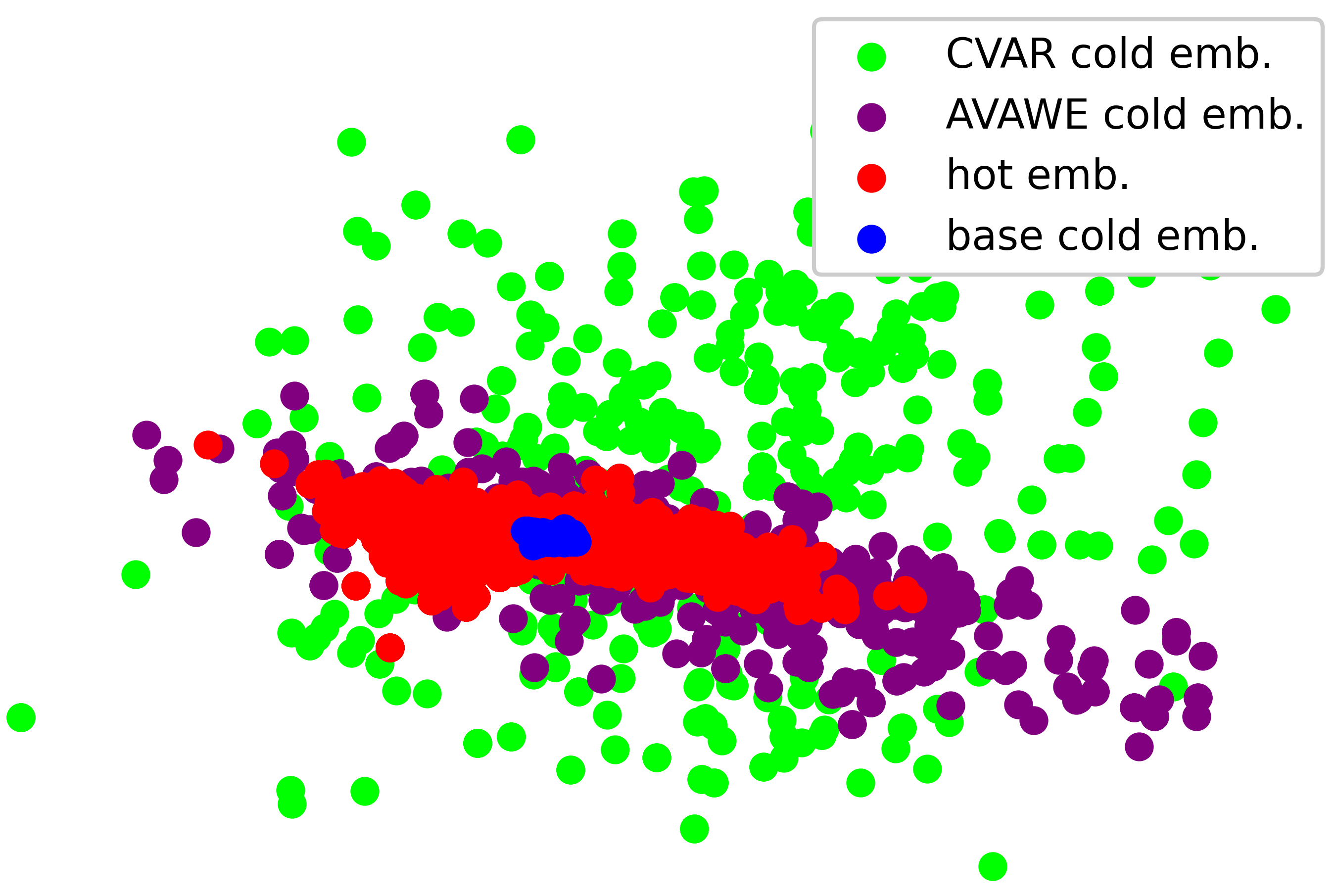}
    \caption{Distribution of the cold items' ID embedding generated by different methods as well as the item ID embedding of the hot items.}
    \label{fig:item_emb}\
    \vspace{-0.5cm}
\end{figure}

Figure~\ref{fig:item_emb} illustrates the cold item embedding generated by different methods as well as the embedding of hot items in the pre-trained base model. We use DeepFM as the backbone model and PCA for dimension reduction. The cold item embeddings of the base model, which is only updated by limited interaction, are not well trained and thus near the initialized points. While the distribution of cold item embeddings generated by the CVAR are quite different from the warm item embedding. With alignment by the adversarial module, our method AVAEW can properly warm-up item embeddings for the cold items to increase their recommendation performance.

Then we further demonstrate the compatibility of AVAEW on different backbones. Table~\ref{TAB:DiffBackbones} shows the AUC performance of AVAEW with different types of backbones recommendation models on MovieLens-1M, MovieLens-25M, and Taobao-Ad, respectively. The results show that AVAEW can boost all kinds of tested backbones in cold item recommendations. In the two large datasets (MovieLens-25M and Taobao-AD), the AVAWE can increase the AUC by over 0.1 in the cold item recommendation, which is a huge improvement in the practical large-scale recommender system.

\subsection{Online Validation}
We further validate the proposed AVAEW on a large-scale real-world news recommendation system, which serves over 100 million users and receives over one hundred thousand new content each day. Here are usually four stages in the large-scale recommender system~\cite{wang2020cold}, namely matching, pre-ranking, ranking, and re-ranking. We applied AVAEW in the pre-rank stage, which is designed to predict the CTR of items and filter out plenty of items with the lowest CTR scores. We choose to apply AVAEW in the pre-rank stage because we found out that most of the new items are ranked relatively low and filtered out. 
The backbone model of the pre-rank stage is DSSM~\cite{huang2013learning}. There are 74 features for the backbone model, which include 49 user features and 25 item features, among which we use 11 important item side information for the inputs of AVAEW.

In the online A/B test, we warm up the cold video-type items and focus on the following metrics: exposure rate, and video view. 
The exposure rate is the number of unique recommended items. A higher exposure rate indicates that a cold item has more chances to be accessed by users and the diversity of recommended content is increased.
The video view is the number of items viewed by users, reflecting the satisfaction of users with the content.

We run the online A/B test for 8 days, which involves over 80 thousand users for each of the base model and AVAEW model experiments. Compared with the base model, AVAEW increases the exposure rate by $+1.6019\%$ and the video views by $+1.1396\%$. In order to measure the satisfaction of users on the recommended content more accurately, we further measure the deep video views as the views that the time length of video play $>$ 30s or finish rate $>$ 0.8, where finish rate is the ratio of the time length of video play to the video length. By doing so, we exclude the views that users may just be attached to by titles but not the content. For the deep video view, AVAEW achieves $+0.3352\%$ improvement on long videos (video length $>=$ 30s) and $+0.7268\%$ improvement on short videos (video length $<$ 30s). 





\section{Conclusion and Discussion}
In this paper, we proposed a novel method, AVAEW, for alleviating the item ID embedding gap between cold and other items. By using an adversarial module to address the item ID embedding gap problem, the AVAEW generates warm-up item ID embeddings that are more suited to the recommender system. Extensive experiments on public datasets validated the effectiveness of the proposed method by comparing AVAEW with the SOTA methods in three main categories of item cold recommendation methods. Moreover, we demonstrated the compatibility of AVAEW on different kinds of backbone models. 

We also showed that AVAEW can be easily applied in the real-world large-scale recommendation system and improve item cold start without extra development of data construction or training pipeline. The model is capable of continuously adapting to the new items. Our method also has the potential to be used in user cold start, because like the item side information, the user profile, such as age and occupation, can partially reflect the user interest and thus can facilitate the user embedding learning. However, the user profile data is privacy sensitive, and the item cold start problem is more serious in our online platform. Thus we first focus on the item cold start in this study.

\bibliographystyle{splncs04}
\bibliography{sample-base}

\end{document}


%
\setcounter{tocdepth}{2}
\tableofcontents

\section{Detailed Introduction of Related Work}
The existing methods proposed to solve the item cold start problem mainly fall into three categories. The methods in the first category attempt to promote the robustness of the recommendation model in absence of item ID embedding. For example, DropoutNet uses dropout on item ID embedding in the model training~\cite{volkovs2017dropoutnet}. By doing so, the model can be generalized to the inference stage where the item ID embedding is not well-trained. Some other methods achieve the same goal by randomly dropout the user-item interaction~\cite{zheng2021multi} or using attention mechanism~\cite{shiattention2018}.

The methods in the second category focus on improving learning efficiency with a limited amount of interaction data. For example, Meta embedding ~\cite{pan2019warm} learns the ID embedding of new items with meta-learning for quick updates during the first few batches. CMML~\cite{fengcmml2021} further explicitly model the task context representation in the meta-learning framework for facilitating fast adaption.  On the other hand, FORM~\cite{sunform2021} addresses the cold-start problem from the perspective of online learning. 

The methods in the third category aim to leverage the side information of items to facilitate the initialization of the item ID embedding.
For example, MWUF~\cite{zhu2021learning} proposes a meta Scaling/Shifting module to map the cold item ID embedding to warm-up item ID embedding, which is more close the ID embedding of warm items and suits the recommendation system. To utilize the side information as the alternative of the item embedding, \cite{zhu2020recommendation} alternating feed the item embedding and side representation to the model. \cite{wei2021contrastive} use contrastive learning to fuse the item ID embedding and the side information representation. Particularly, GAR~\cite{chen2022generative} designs an adversarial training strategy: train an item generator via maximizing the ranking scores of cold items and train the recommendation model to decrease the corresponding ranking scores. However, they do not directly
consider the consistency of the distribution, nor do they strictly guarantee to
reduce the distribution gap during the warm-up process, which reduces their
recommendation effect.

\section{Detail of Datasets used in Offline Experiments}
\begin{enumerate}
    \item MovieLens-1M~\footnote{\url{https://grouplens.org/datasets/movielens/1m/}}: We first evaluate the methods on a relatively small size MovieLens-1M dataset, one of the most well-known benchmark data, to simulate the scenarios where a limited amount of data is available. The data consists of 1 million movie ranking data over 4000 movies and 6000 users. The side information of each movie includes its title, year of release, and genre. Other features include the user’s ID, age, gender, and occupation.
    \item MovieLens-25M~\footnote{\url{https://grouplens.org/datasets/movielens/25m/}}: 
    Furthermore, we evaluate the methods on a large version of MovieLens dataset: MovieLens-25M, which contains 25000095 ratings and 1093360 tag applications across 62423 movies. These data were created by 162541 users between Jan 09, 1995 and Nov 21, 2019. The side information of each movie includes its title, year of publish, genres, and top10 tags. Other features include the user id and timestamp.
    \item Taobao Display Ad Click~\footnote{\url{https://tianchi.aliyun.com/dataset/dataDetail?dataId=56}}: Taobao Display Ad Click is a dataset of click rate prediction about display Ad, which is displayed on the website of Taobao. 1140000 users are randomly sampled from the website of Taobao for 8 days of ad display/click logs to form the original sample skeleton, which contains 26 million records. There are 5 side information features about an item (Ad) and 11 other features.
\end{enumerate}

The statistic of the used datasets in the experiments are summarized in Table~\ref{tab:data_statistic}
\begin{table}[h]
    \centering
    \begin{tabular}{c|c|c|c|c}
         \hline
         data name & users \#& items \# & ratings \# & \begin{tabular}[x]{@{}c@{}}item/other\\feature \#\end{tabular}
            \\
         \hline
         \hline
         MovieLen-1M   & 6,040   & 3,900 & 1,000,209 & 5/6 \\
         MovieLen-25M  & 162,541  & 62,423 & 2,5000,095 & 5/2 \\
         Taobao-AD     & 1,140,000 & 1,433 & 26 million & 6/9 \\
         \hline
    \end{tabular}
    \caption{Statistic of used datasets in the experiments}
    \label{tab:data_statistic}
\end{table}

\section{Ablation Study}
Tabel~\ref{TAB:abalation} shows the ablation result in the effectiveness of different loss terms in two large datasets. Overall, each loss term contributes to the performance of AVAEW.

\begin{table}[h]
\centering
\begin{tabular}{|l|c|c|c|c|}
\hline
\multirow{2}{*}{}       & \multicolumn{2}{l|}{Taobao AD}         & \multicolumn{2}{l|}{MovieLens -25M}    \\ \cline{2-5} 
                        & {DeepFM}  & IPNN    & {DeepFM}  & IPNN    \\ \hline
w/o $\mathcal{L}_{R}$ &{-0.81\%} & 0.08\%  & {-0.37\%} & -0.27\% \\ \hline
w/o $\mathcal{L}_{WD}$ & {-0.70\%} & 0.10\%  & {0.06\%}  & -0.60\% \\ \hline
w/o $\mathcal{L}_{GD}$  & {-0.74\%} & -1.11\% & {-0.03\%} & -0.52\% \\ \hline
w/o $\mathcal{L}^{'}_{GD}$ & {-0.72\%} & -0.02\% & {-0.05\%} & 0.08\%  \\ \hline
\end{tabular}
\caption{Performance change in warm-c phase: $(c- b)/b$, $c$: AUC of AVAEW w/o ($\mathcal{L}_{R}$, $\mathcal{L}_{WD}$, $\mathcal{L}_{GD}$, $\mathcal{L}^{'}_{GD}$) , $b$: AUC of AVAEW}
\label{TAB:abalation}
\vspace{-5mm}
\end{table}

\bibliographystyle{splncs04}
\bibliography{sample-base}